\documentclass[twocolumn,pre,superscriptaddress,floatfix,amsmath]{revtex4}
\usepackage{graphicx}

\newcommand{\myav}[1]{\langle{#1}\rangle}
\newcommand{\mymat}[1]{{\mathbf #1}}
\newcommand{\myorder}[1]{\{#1\}}
\newcommand{\zero}{\myorder{0}}
\newcommand{\one}{\myorder{1}}
\newcommand{\mysum}[1]{\underset{\scriptstyle\alpha\in #1}{\textstyle\sum}}
\newcommand{\sumzero}{\mysum\zero}
\newcommand{\sumone}{\mysum\one}
\newcommand{\torate}[1]{\overset{#1}{\to}}
\newcommand{\tofromrate}[2]{\underset{#2}{\overset{#1}{\rightleftharpoons}}}
\newcommand{\latin}[1]{\emph{#1}}
\newcommand{\species}[1]{\emph{#1}}

\begin{document}

\title{Exact results for noise power spectra in linear biochemical
reaction networks}

\author{Patrick B. Warren}
\affiliation{Unilever R\&D Port Sunlight, Bebington, Wirral, CH63 3JW, UK.}
\author{Sorin {T\u anasa-Nicola}}
\affiliation{FOM Institute for Atomic and Molecular Physics,
Kruislaan 407, 1098 SJ Amsterdam, The Netherlands.}
\author{Pieter Rein ten Wolde}
\affiliation{FOM Institute for Atomic and Molecular Physics,
Kruislaan 407, 1098 SJ Amsterdam, The Netherlands.}

\date{December 2005}

\begin{abstract}
We present a simple method for determining the exact noise power
spectra in linear chemical reaction networks.  We apply the method to
networks which are representative of biochemical processes such as
gene expression and signal detection.  Our results clarify how noise
is transmitted by signal detection motifs, and indicate how to
coarse-grain networks by the elimination of fast reactions.
\end{abstract}

\maketitle

\section{Introduction}
\label{sec:intro}
It is now clear that chemical noise can sometimes play a significant
role in the functioning of biochemical reaction networks \cite{RWA}.
Such noise is implicated in the spontaneous flipping of genetic
switches \cite{ARMcA}, and has been shown to have an adverse effect on
the functioning of a synthetic chemical oscillator \cite{Elowitz}.  In
some circumstances it has also been argued that noise can have a
beneficial effect, for example in stochastic focusing \cite{PBE}.

Chemical noise in reaction networks can be characterised by the root
mean square (rms) deviation of number of molecules from the mean.
Noise is important if the rms deviation is a significant fraction of
the mean.  This is typically associated with systems where the mean
number of molecules is rather small.  Such situations are
characteristic of gene regulatory networks where the concentrations of
regulatory proteins can be small enough for there to be only a few
molecules in the cell volume.  For example, for the lac operon in
\species{E. coli}, the lac repressor is active at concentrations where
there are only 10--20 molecules present in the cell volume
\cite{SanMAc}.

The mean and the rms deviation are `point statistics', and have
frequently been used to characterise the stochastic properties of
biochemical reaction networks.  In the present paper, we focus on the
noise power spectra, which are a more refined characterisation of the
stochastic properties.  We also focus on linear reaction networks for
which the noise power spectra and point statistics can be calculated
exactly.  In the next section, we develop a general theory for
computing the noise power spectra in linear reaction networks.  We
relegate technical details to the Appendix.  In the following sections
we apply the general theory to increasingly elaborate networks which
model basic processes of biochemical interest, such as gene expression
and signal detection.

\section{General theory}
\label{sec:gen}
We define a \emph{linear reaction network} to be a network of chemical
reactions which does not involve bimolecular or higher order reactions.
For such a network, the chemical rate equations can be written as
\begin{equation}
\frac{d N_i}{dt} = {\textstyle\sum_j}K_{ij} N_j + b_i\label{eq:cre}
\end{equation}
where $N_i$ is the number of molecules of species $i$, $K_{ij}$ is a
matrix of rate coefficients, and $b_i$ are source terms.
Eq.~\eqref{eq:cre} is a set of linear ordinary differential equations,
which explains the origin of the phrase `linear reaction network'.
Appendix \ref{sec:appkb} indicates how $K_{ij}$ and $b_i$ are related
to the stoichiometry matrix and reaction rates which specify the
network.

Let $N_i(t)$ be the instaneous number of molecules of species $i$ in
the system.  Define the mean value of $N_i$ in steady state to be
$\myav{N_i}$, and the mean value out of steady state to be
$\myav{N_i}_t$.  By taking moments of the chemical master equation
(see Appendix), one can prove that
\begin{equation}
\frac{d\myav{N_i}_t}{dt} = {\textstyle\sum_j}K_{ij}\myav{N_i}_t 
+ b_i.\label{eq:nit} 
\end{equation}
Therefore, the chemical rate equations are \emph{exact} for a linear
reaction network, provided we work with the mean quantities
$\myav{N_i}_t$.  In steady state, the mean values therefore solve
\begin{equation}
{\textstyle\sum_j}K_{ij}\myav{N_i} + b_i = 0.\label{eq:ni}
\end{equation}
This is a system of linear simultaneous equations which can be readily
solved. 

We now define the deviation away from the steady state mean values to be
\begin{equation}
\Delta N_i(t) = N_i(t) - \myav{N_i}
\end{equation}
and introduce the steady-state variance-covariance matrix
\begin{equation}
S_{ij}=\myav{\Delta N_i\,\Delta N_j}.
\end{equation}
The diagonal elements of this are the variances
\begin{equation}
\sigma_i^2=S_{ii}=\myav{\Delta N_i^2}.
\end{equation}
The rms deviation mentioned in the introduction is $\sigma_i$.  It is
often the case that $\sigma_i^2\sim\myav{N_i}$ and in fact some
authors define the `noise strength' to be $\sigma_i^2/\myav{N_i}$
although we will not explicitly use this in the present paper.  This
scaling means that the relative rms deviation $\sigma_i/\myav{N_i}\sim
1/\surd\myav{N_i}$ which indicates that the relative important of
noise decreases inversely with the square root of the number of
molecules involved.  It is the basic reason why noise is important for
systems where the number of molecules is small.

As many authors have noticed \cite{Berg, McAA, TvO}, the
variance-covariance matrix at steady state can also be obtained by
taking moments of the chemical master equation.  The details are
unimportant for our arguments, and are relegated to Appendix
\ref{sec:appkb}.  For the remainder of the main text, we assume that the
means and variance-covariance matrix have been computed.  We will say
that these comprise the \emph{point statistics} of the network.

To refine the description of the steady state beyond the point
statistics, we introduce the set of correlation functions
\begin{equation}
C_{ij}(t)=\myav{\Delta N_i(0)\,\Delta N_j(t)}.
\end{equation}
In steady state, the value chosen for the time origin is unimportant
(therefore the average could be regarded as a time average over
starting times, instead of an ensemble average).  The correlation
functions have the properties
\begin{equation}
\begin{array}{l}
C_{ij}(0) \to S_{ij}\quad\text{as}\quad t\to0,\\[6pt]
C_{ij}(t) \to 0\quad\text{as}\quad t\to\infty,\\[6pt]
C_{ij}(t) = C_{ji}(-t)
\end{array}
\end{equation}
(the first of these is not always true; see section \ref{sec:fast}).
The last property shows that the autocorrelation functions are
time-symmetric, $C_{ii}(t) = C_{ii}(-t)$, but it does not follow that
the cross-correlation functions are time-symmetric, as is illustrated
by the example in section \ref{sec:gem}.

Closely related to the autocorrelation functions are the noise power
spectra, which are the central topic of study in the present paper.
These are defined to be
\begin{equation}
P_i(\omega) = \myav{|\Delta N_i(\omega)|^2} 
= 2{\int_0^\infty}\!\!dt\,
\cos\omega t \,C_{ii}(t)\label{eq:power}
\end{equation}
where $\Delta N_i(\omega)$ is the Fourier transform of $\Delta
N_i(t)$.  We shall use the second form in Eq.~\eqref{eq:power} to
compute the noise power spectra.  

Usually the noise power spectra obey the sum rule (for an exception,
see section \ref{sec:fast}),
\begin{equation}
\sigma_i^2 = \frac{1}{2\pi} {\int_0^\infty}\!\!
d\omega\,P_i(\omega).\label{eq:sumrule}
\end{equation}

Now we turn to the computation of the correlation functions and the
noise power spectra.  In Appendix \ref{sec:appc} we prove that the
correlation functions satisfy
\begin{equation}
\frac{dC_{ki}}{dt} = {\textstyle\sum_j}K_{ij}C_{kj}.\label{eq:corr} 
\end{equation}
If we subtract Eq.~\eqref{eq:ni} from Eq.~\eqref{eq:nit}, we find that
the mean deviation out of steady state obeys
\begin{equation}
\frac{d\myav{\Delta N_i}_t}{dt} 
= {\textstyle\sum_j}K_{ij}\myav{\Delta N_i}_t.\label{eq:dnit} 
\end{equation}
This shows that the correlation functions decay in exactly the same
way as deviations away from steady state, so that Eq.~\eqref{eq:corr}
is an example of a regression theorem \cite{gardinerbook}.

Eq.~\eqref{eq:corr} is to be solved with the initial conditions
$C_{ij}(0) = S_{ij}$.  A convenient approach is by the use of
the Laplace transform, which allows the initial conditions to be
automatically incorporated.  Taking the Laplace transform of
Eq.~\eqref{eq:corr} shows that
\begin{equation}
s \tilde C_{ki} - S_{ki} 
= {\textstyle\sum_j}K_{ij}\tilde C_{kj}\label{eq:lcorr}
\end{equation}
where
\begin{equation}
\tilde C_{ij}(s) = {\int_0^\infty}\!\!dt\,e^{-st}
C_{ij}(t).\label{eq:laplace}
\end{equation}
Eq.~\eqref{eq:lcorr} is a set of linear simultaneous equations which
can readily be solved for $\tilde C_{ij}(s)$.

The power spectra now follow almost for free.  Comparing
Eq.~\eqref{eq:laplace} with Eq.~\eqref{eq:power} shows that
\begin{equation}
P_i(\omega) = \tilde C_{ii}(i\omega) + \tilde C_{ii}(-i\omega)
,\label{eq:l2p}
\end{equation}
in other words $P_i(\omega)$ is twice the real part of the analytic
continuation of $\tilde C_{ii}(s)$ to $s = i\omega$.

Eqs.~\eqref{eq:lcorr} and \eqref{eq:l2p} are the key results of this
section.  They indicate how exact results for the power spectra can be
obtained for an arbitrary linear reaction network by purely algebraic
methods.  In Appendix \ref{sec:appl} we prove that the same results
can be obtained from the chemical Langevin equations
described by Gillespie \cite{GillCLE}.

It is interesting to note that the power spectra are simpler to obtain
than the correlation functions themselves, which require an inverse
Fourier or Laplace transform.  To undertake such a step requires in
general that a polynomial be factorised, which is not always possible.
In fact the route to the power spectra, via Eqs~\eqref{eq:ni},
\eqref{eq:lcorr} and \eqref{eq:app:vcm} in the Appendix, just involves
the solution of linear simultaneous equations.

\begin{figure}
\begin{center}
\includegraphics{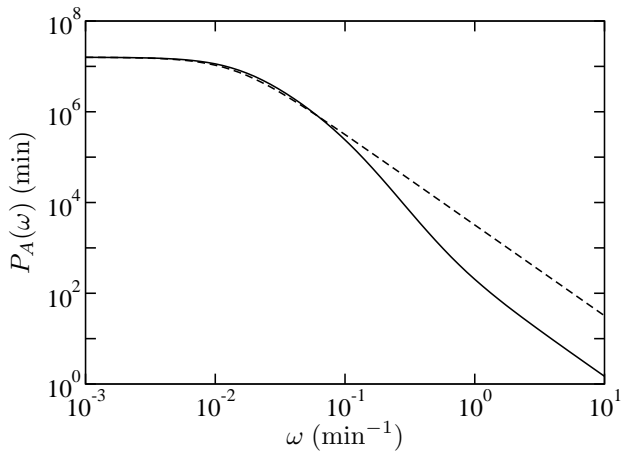}
\end{center}
\caption[?]{Power spectra for the gene expression models in
Eq.~\eqref{eq:A} (Eq.~\eqref{eq:pnk}; dashed line) and
Eq.~\eqref{eq:MA} (Eq.~\eqref{eq:MApa}; solid line).  Parameters for
the explicit mRNA model of Eq.~\eqref{eq:MA} are $k = 2.76\,
\mathrm{min}^{-1}$, $\lambda = 0.12\, \mathrm{min}^{-1}$, $\rho =
3.2\, \mathrm{min}^{-1}$, and $\gamma = 0.016\, \mathrm{min}^{-1}$.
These correspond to cro in a recent model of phage lambda
\cite{SanMac2}.  Parameters for the implicit mRNA model of
Eq.~\eqref{eq:A} are $k = 1.36\, \mathrm{min}^{-1}$, $n = 48.1$,
$\gamma = 0.0142\, \mathrm{min}^{-1}$.  These are chosen so that the
two models have matched point statistics, $\myav{N_A} = 4.6\times10^3$
and $\sigma_A / \myav{N_A} = 0.07$, and matching values of $P_A(0) =
1.6\times10^7\,\mathrm{min}$.\label{fig:fig1}}
\end{figure}

\section{Applications}
\subsection{Gene expression models}
\label{sec:gem}
We first consider a very simple model of gene expression
\begin{equation}
\torate{k}n\,\text{A},\qquad \text{A}\torate{\gamma}
\label{eq:A}
\end{equation}
which represents a birth-death process in which $n$
copies of A are generated each time the first (birth) reaction fires.
This represents the `burstiness' of gene expression in prokaryotes
\cite{TvO}.  The point statistics are
\begin{equation}
\myav{N_A}=\frac{nk}{\gamma},\qquad\sigma_A^2=\frac{n+1}{2}\myav{N_A}
\label{eq:nk}
\end{equation}
The chemical rate equation is $dN_A/dt = nk - \gamma N_A$ so the 
autocorrelation function obeys $dC_{AA}/dt = - \gamma C_{AA}$.  The
solution, and its Laplace transform, are 
\begin{equation}
C_{AA}(t)=\sigma_A^2 e^{-\gamma t},\quad
\tilde C_{AA}(s)=\frac{\sigma_A^2}{s+\gamma}.
\end{equation}
The corresponding power spectrum is
\begin{equation}
P_A(\omega)=\frac{2\gamma\sigma_A^2}{\omega^2+\gamma^2}.\label{eq:pnk}
\end{equation}

Now we turn to a more realistic representation of gene expression
which explicitly includes mRNA, namely
\begin{equation}
\torate{k}\text{M},\quad \text{M}\torate{\lambda},
\quad \text{M}\torate{\rho}\text{M}+\text{A},\quad
\text{A}\torate{\gamma}.\label{eq:MA}
\end{equation}
The first two reactions represent generation and degradation of M
which corresponds to mRNA.  The second two reactions represent
generation of the protein A from M and the degradation of A.  This is
a model which has been analysed by other workers \cite{Berg, McAA,
TvO}, and we recover previously known results for the point statistics.

The mean values in steady state are
\begin{equation}
\myav{N_M}=\frac{k}{\lambda},\quad
\myav{N_A}=\frac{\rho}{\gamma}\myav{N_M}=\frac{bk}{\gamma}.
\label{eq:bk}
\end{equation}
In this, $b=\rho/\lambda$ is the mean number of copies of A produced
in the lifetime of one mRNA and can be taken to represent the
`burstiness' of gene expression.  Comparing the last of
Eq.~\eqref{eq:bk} with the first of Eq.~\eqref{eq:nk}, it appears that
$b$ in the present model corresponds to $n$ in the previous implicit
mRNA model.  As we shall see though, this interpretation does not
carry through to the other statistical properties.

The steady state variances and covariance are 
(see Appendix \ref{sec:appkb}) 
\begin{equation}
\begin{array}{l}
\displaystyle
\sigma_M^2=\myav{N_M},\quad
\sigma_A^2=\Bigl(1+\frac{\rho}{\gamma+\lambda}\Bigr)\,\myav{N_A},\\[18pt]
\displaystyle
S_{M\!A}=\frac{\rho\myav{N_M}}{\gamma+\lambda}
=\frac{\gamma\myav{N_A}}{\gamma+\lambda}.\label{eq:MAvcm}
\end{array}
\end{equation}
In this case, comparing Eq.~\eqref{eq:MAvcm} with Eq.~\eqref{eq:nk},
we see that the correspondence $b\leftrightarrow n$ fails for
$\sigma_A^2$.

Now we turn to the computation of the correlation functions and the
power spectra.  The chemical rate equations are 
\begin{equation}
\frac{d N_M}{dt} = k - \lambda N_M,\quad
\frac{d N_A}{dt} = \rho N_M - \gamma N_A.
\end{equation}
The correlation functions obey the homogeneous form of these
equations.  Taking the Laplace transform as indicated in the previous
section, we find
\begin{equation}
\begin{array}{l}
s\tilde C_{M\!M} - \sigma_M^2 = - \lambda\tilde C_{M\!M},\\[6pt]
s\tilde C_{AM} - S_{AM} = - \lambda\tilde C_{AM},\\[6pt]
s\tilde C_{M\!A} - S_{M\!A} = \rho \tilde C_{M\!M}
 - \gamma \tilde C_{M\!A},\\[6pt]
s\tilde C_{AA} - \sigma_A^2 = \rho \tilde C_{AM} - \gamma \tilde C_{AA}.
\end{array}
\end{equation}
Because $C_{M\!A}(t)\ne C_{AM}(t)$, there are four different
Laplace-transformed correlation functions (note though that
$S_{M\!A}=S_{AM}$).  The solutions are
\begin{equation}
\begin{array}{l}
\displaystyle
\tilde C_{M\!M}=\frac{\sigma_M^2}{s+\lambda},\qquad
\tilde C_{AM}=\frac{S_{AM}}{s+\lambda},\\[12pt]
\displaystyle
\tilde C_{M\!A}=\frac{S_{M\!A}}{s+\gamma}+
\frac{\rho\sigma_M^2}{(s+\lambda)(s+\gamma)},\\[12pt]
\displaystyle
\tilde C_{AA}=\frac{\sigma_A^2}{s+\gamma}+
\frac{\rho S_{AM}}{(s+\lambda)(s+\gamma)}.
\end{array}
\end{equation}
In this particular case, the inverse Laplace transforms can easily be
found by the method of partial fractions.  We give only the result
for the cross-correlation function, which exhibits the interesting time
asymmetry mentioned above,
\begin{equation}
C_{M\!A}(t)=
\left\{
\begin{array}{ll}
\displaystyle 
S_{M\!A} \,e^{-\gamma t}+
\rho\sigma_M^2\,
\frac{e^{-\gamma t}-e^{-\lambda t}}{\lambda-\gamma}
& (t\ge0)\\
\displaystyle 
S_{M\!A} \,e^{-\lambda|t|} 
\phantom{\frac{e^{-\gamma t}-e^{-\lambda t}}{\lambda-\gamma}}
& (t\le0)
\end{array}
\right.
\end{equation}
(using $C_{M\!A}(-t)=C_{AM}(t)$ for the second line).

Finally, the power spectra are obtained.  The power spectrum for M is
$P_M(\omega) = {2\lambda\myav{N_M}}/{(\omega^2+\lambda^2)}$.  This
might have been anticipated as a special case of the previous model
with $n=1$, since as far as M is concerned it is undergoing a simple
birth-death process.  The power spectrum for A is
\begin{equation}
P_A(\omega)=\frac{2\gamma\myav{N_A}(\omega^2+\lambda(\lambda+\rho))}%
{(\omega^2+\lambda^2)(\omega^2+\gamma^2)}.\label{eq:MApa}
\end{equation}
With a suitable choice of $n$ and rate cofficients $k$ and $\mu$, the
point statistics of the previous model in Eq.~\eqref{eq:A} could be
matched to the present model in Eq.~\eqref{eq:MA}.  However, it is in
general impossible to match the power spectra since the $\omega$
dependence is different.  Fig.~\ref{fig:fig1} shows the power spectra
for the two models with matched point statistics.  The presence of two
correlation times in the explicit mRNA model compared to a single
relaxation time for the implicit mRNA model is clearly seen.

We have gone through the calculations in some detail for these two
model, but for the subsequent calculations we will only give the final
results.

\subsection{Signal detection}
A very simple model of signal detection is the reaction
\begin{equation}
\text{A}\torate{\nu}\text{B},\quad\quad 
\text{B}\torate{\mu}\label{eq:AB}
\end{equation}
where A is the input signal and B is the output signal.
Eq.~\eqref{eq:AB} has a chemical rate equation
\begin{equation}
\frac{d N_B}{dt}=\nu N_A-\mu N_B+\eta.\label{eq:ABcle}
\end{equation}
We have included an additional noise term $\eta$ in this, so it is
actually a chemical Langevin equation.  Taking the Fourier transform
yields
\begin{equation}
i\omega N_B=\nu N_A-\mu N_B + \eta,
\quad\text{or}\quad
N_B=\frac{\nu N_A+\eta}{i\omega+\mu}.
\end{equation}
The mean square modulus is then
\begin{equation}
\myav{|N_B|^2}=\frac{\nu^2\myav{|N_A|^2}+\myav{|\eta|^2}}
{\omega^2+\mu^2}.
\end{equation}
We have assumed that $\eta$ is uncorrelated with $N_A$.  The theory of
chemical Langevin equations \cite{GillCLE} (see also Appendix
\ref{sec:appl}) shows that $\myav{|\eta|^2} = 2\mu\myav{N_B}$, and for
$\omega>0$ one has $\myav{|N_i|^2} = \myav{|\Delta N_i|^2}$, thus
finally
\begin{equation}
P_B(\omega)=\frac{\nu^2}{\omega^2+\mu^2}\, P_A(\omega)
+\frac{2\mu\myav{N_B}}{\omega^2+\mu^2}.
\label{eq:psf}
\end{equation}
This result has quite a striking interpretation.  The total noise in
the output signal is made up of an \emph{extrinsic} contribution
(first term) plus an \emph{intrinsic} contribution (second term).
Moreover, the extrinsic noise is equal the input signal noise
multiplied by a low-pass filter function.  This is analogous to the
behaviour of a passive RC circuit element \cite{SAR}.  This result, in
one form or another, was derived by Paulsson \cite{Paulsson}, and by
Shibata and Fujimoto \cite{Shibata}.  We refer to it as the `spectral
addition rule'.  It is a potentially important result because it
indicates that there is a trade-off between signal amplification by
the detection motif and signal contamination by added intrinsic noise.

Our exact results for linear reaction networks enable us to test the
spectral addition rule.  We find that its validity is limited by the
assumption that the intrinsic noise $\eta$ is uncorrelated with the
input signal $N_A$.  In particular, if the detection motif consumes
molecules of the input signal, a correlation can arise which spoils
the spectral addition rule.

To demonstrate this, we now consider the effect of adjoining
Eq.~\eqref{eq:AB} onto the gene expression models in the previous
section.  We present results for the explicit mRNA model, but similar
conclusions can be drawn for the implicit mRNA model too \cite{TNWtW}.
We therefore solve
\begin{equation}
\torate{k}\text{M}\torate{\lambda},
\quad \text{M}\torate{\rho}\text{M}+\text{A},\quad
\text{A}\torate{\gamma},\quad
\text{A}\torate{\nu}\text{B}\torate{\mu}.
\label{eq:MAB}
\end{equation}
We find that
\begin{equation}
\frac{P_A(\omega)}{\myav{N_A}}=
\frac{2(\gamma+\nu)(\omega^2+\lambda(\lambda+\rho))}
{(\omega^2+\lambda^2)(\omega^2+(\gamma+\nu)^2)}
\end{equation}
and
\begin{equation}
\frac{P_B(\omega)}{\myav{N_B}}=
\frac{2\mu[(\omega^2+\lambda^2)(\omega^2+(\gamma+\nu)^2)
+\lambda\nu\rho(\gamma+\nu)]}
{(\omega^2+\lambda^2)(\omega^2+(\gamma+\nu)^2)(\omega^2+\mu^2)}
\end{equation}
with $\nu\myav{N_A}=\mu\myav{N_B}$.  It is not hard to demonstrate
that the spectral addition rule \emph{fails} for these expressions.

As an aside, one can show $\sigma_B^2/\myav{N_B} <
\sigma_A^2/\myav{N_A}$.  Since the $\text{A}\to\text{B}$ reaction in
Eq.~\eqref{eq:MAB} can be regarded as post-translational modification
step, this shows that post-translational modification can reduce the
noise associated with gene expression.  The reason is that the
post-translational modification reaction smoothes the `burstiness' of
gene expression by acting as a low-pass filter.

To appreciate the influence of coupling the detection reaction to the
input signal noise, we now change the detection scheme so that the
signal molecules A are \emph{not} consumed, by replacing
Eq.~\eqref{eq:AB} with
\begin{equation}
\text{A}\torate{\nu}\text{A}+\text{B},\quad\quad 
\text{B}\torate{\mu}.\label{eq:AAB}
\end{equation}
As far as B is concerned, it is important to note that the \emph{same}
chemical Langevin equation Eq.~\eqref{eq:ABcle} holds for this
detection scheme as for the previous scheme.  For this variant we find
$P_A(\omega)$ is as given in Eq.~\eqref{eq:MApa} and
\begin{equation}
\frac{P_B(\omega)}{\myav{N_B}}=
\frac{2\mu[(\omega^2+\lambda^2)(\omega^2+\gamma(\gamma+\nu))
+\gamma\nu\rho\lambda]}
{(\omega^2+\gamma^2)(\omega^2+\lambda^2)(\omega^2+\mu^2)}
\label{eq:pBMAB}
\end{equation}
One can check that in this case $P_A(\omega)$ and $P_B(\omega)$ do
obey the spectral addition rule.

Whilst we have only demonstrated the failure of the spectral addition rule
for one particular case, the result is suggestive of the general
conclusion that the spectral addition rule will only hold for detection
schemes which do not consume input signal molecules.  These
conclusions can also be reached by analysing the chemical Langevin
equations for the whole network \cite{TNWtW}.

\subsection{Fast reactions}
\label{sec:fast}
There is a growing literature on the adiabatic elimination of fast
reactions for stochastic chemical kinetics \cite{Kepler01,
Haseltine02, Shibata03a, Shibata03, Bundschuh, Rao03, Roussel04,
Chatterjee05, Cao05, Straube05}.  For example Kepler and Elston
\cite{Kepler01}, and Shibata \cite{Shibata03a, Shibata03}, have a
formalism that permits a systematic approach to the problem, involving
the identification of the fast and slow variables in the system.  In a
similar example, Bundschuh \latin{et al} present simulation results
which support the general strategy of elimination of fast variables in
terms of slow variables \cite{Bundschuh}.  Here we examine the
procedure for elimination of a fast equilibration reaction in the
context of the \emph{exact} results for a linear reaction network.
Our results shed light on the way in which fast equilibration
reactions can be systematically eliminated from a reaction network,
whilst preserving the noise attributes.

To make a concrete example, we suppose that the signal detection motif
of the previous section consists of a fast
binding-unbinding step, followed by a slower detection step.  We
replace $\text{A}\to\text{B}$ by
$\text{A}\rightleftharpoons\text{A}^*\to\text{B}$ where A$^*$
represents the bound state.  We suppose that the substrate which binds
A is in excess, so that our reaction network is still a linear
network.  The reaction scheme in its totality is now
\begin{equation}
\begin{array}{ll}
\torate{k}\text{M}\torate{\lambda},
&{}\quad \text{M}\torate{\rho}\text{M}+\text{A},\\[6pt]
\text{A}\torate{\gamma},
&{}\quad
\text{A}\tofromrate{k_{\!f}}{k_b}\text{A}^*\torate{\nu}\text{B}\torate{\mu}.
\end{array}
\label{eq:MAAsB}
\end{equation}
The noise statistics for this network can be solved but the
expressions are rather lengthy.  We therefore focus on the limit of a
fast binding-unbinding reaction.  

The equilibrium constant $K$ for the binding-unbinding reaction is
defined via
\begin{equation}
{k_{\!f}}=K\,{k_b}.
\end{equation}
To take the limit of fast equilibration, we keep $K$ finite and allow
$k_b\to\infty$.  The results are as follows.  Let us first consider
the final product species B.  We find $P_B(\omega)$ is given by
Eq.~\eqref{eq:pBMAB} but with `renormalised' reaction rates
\begin{equation}
\gamma_R=\frac{\gamma}{K+1},\quad\nu_R=\frac{K\nu}{K+1}
\label{eq:renorm}
\end{equation}
The same is true of the variance $\sigma_B^2$.  

We can rationalise this as follows.  The chemical rate equations which
involve the species in the fast equilibration reaction (A and A$^*$)
are
\begin{equation}
\begin{array}{l}
dN_A/dt = \rho N_M-(\gamma+k_{\!f}) N_A+k_b N_{A^*},\\[6pt]
dN_{A^*}/dt = k_{\!f}N_A-(\nu+k_b)N_{A^*},\\[6pt]
dN_B/dt = \nu N_{A^*}-\mu N_B.
\end{array}
\end{equation}
We now make a linear transformation from $N_A$ and $N_{A^*}$ to new
variables $N_X$ and $N_Y$ :
\begin{equation}
\begin{array}{l}
N_X = N_A + N_{A^*},\\[6pt]
N_Y = K N_A - N_{A^*}.
\end{array}
\label{eq:lintran}
\end{equation}
These are chosen because $N_X$ is a \emph{slow} variable which is
conserved by the equilibration reaction, and $N_Y$ is a \emph{fast}
variable which vanishes in steady state.  In terms of these
concentration variables, the chemical rate equations become
\begin{equation}
\begin{array}{l}
\displaystyle
dN_X/dt = \rho N_M - \frac{\gamma+K\nu}{K+1}\,N_X-
\frac{\gamma-\nu}{K+1}\,N_Y,\\[12pt]
\displaystyle
dN_Y/dt +(K+1) k_b N_Y = K\rho N_M-\frac{\gamma-\nu}{K+1}\,N_X,\\[12pt]
\displaystyle
dN_B/dt = \frac{K\nu}{K+1}\,N_X-\frac{\nu}{K+1}\,N_Y-\mu N_B.
\end{array}
\label{eq:cXYB}
\end{equation}
The second of these shows that $N_Y$ relaxes at a rate $\sim k_b$ to a
constant $\sim 1/k_b$, for large $k_b$.  This formalises the
separation of timescales between $N_Y$ and all the other concentration
variables.  In the large $k_b$ limit therefore, $N_Y=0$ and
Eqs.~\eqref{eq:cXYB} become
\begin{equation}
\begin{array}{l}
dN_X/dt = \rho N_M - (\gamma_R+\nu_R) N_X,\\[6pt]
dN_B/dt = \nu_R N_X - \mu N_B.
\end{array}
\end{equation}
with the reaction rates given by Eq.~\eqref{eq:renorm}.  These are (a
subset of) the chemical rate equations for the following reaction
network
\begin{equation}
\torate{k}\text{M}\torate{\lambda},
\quad \text{M}\torate{\rho}\text{M}+\text{X},\quad
\text{X}\torate{\gamma_R},\quad
\text{X}\torate{\nu_R}\text{B}\torate{\mu}.
\label{eq:MXB}
\end{equation}
This is the same as the previously discussed scheme in
Eq.~\eqref{eq:MAB}, with the replacement $\text{A}\leftrightarrow
\text{X}$.  The power spectra follow from Eqs.~\eqref{eq:pBMAB}
accordingly.  The important point to note is that this scheme not only
has the correct chemical rate equations but also has the correct noise
power spectra.

Now let us turn to the power spectra for A and A$^*$ in the original
network.  We find that
\begin{equation}
P_A(\omega) = \frac{1}{(K+1)^2}\,P_X(\omega),\quad
P_{A^*}(\omega) = K^2 P_A(\omega).\label{eq:pAAs}
\end{equation}
Again these results are easy to rationalise, since inverting
Eqs.~\eqref{eq:lintran} and setting $N_Y = 0$ shows that 
\begin{equation}
N_A = \frac{1}{K+1}\,N_X,\quad
N_{A^*} = K\,N_A
\end{equation}
(the proportionality factors become squared when computing the noise
power spectra).

For the variances, $\sigma_A^2$ and $\sigma_{A^*}^2$, we find a
different story though.  In fact, in the limit $k_b\to\infty$, we
find a \emph{breakdown} of the sum rule in Eq.~\eqref{eq:sumrule}.
The reason is that
\begin{equation}
\lim_{k_b\to\infty} \int_0^\infty\!\!d\omega\,
P_i(\omega)\ne
\int_0^\infty\!\!d\omega\,
\lim_{k_b\to\infty} P_i(\omega)
\label{eq:limorder}
\end{equation}
for $i=\text{A}$ and $\text{A}^*$.  The sum rule always holds for the
left hand side of Eq.~\eqref{eq:limorder}, but the limiting power
spectra are defined in terms of the right hand side of
Eq.~\eqref{eq:limorder}.  (As an aside, we find that the sum rule is
satisfied for $P_{A+A^*}(\omega)$, which is equal to $P_X(\omega)$,
confirming that the effect is confined to the individual species in
the fast equilibration reaction.)  To handle this kind of situation,
we define a `sum rule deficit'
\begin{equation}
\Delta\sigma^2_i \equiv \sigma_i^2-\frac{1}{2\pi}\int_0^\infty\!\!d\omega\,
P_i(\omega).
\end{equation}
For the present situation, we find that 
\begin{equation}
\Delta\sigma^2_A = \Delta\sigma^2_{A^*} = \frac{\myav{N_{A^*}}}{K+1}.
\end{equation}
Interestingly, the same result is found for \emph{all} situations
involving a fast equilibration reaction which we have examined.

The origin of the sum rule deficit lies in the noise generated by the
fast equilibration reaction.  Formally, at any finite $k_b$ we can
define the contribution of the equilibration reaction to the total
noise of species $i$ to be
\begin{equation}
\Delta P_i(\omega)=P_i(\omega)-\lim_{k_b\to\infty} P_i(\omega).
\end{equation}
For $i=\text{A}$ and A$^*$, one has that $\Delta P_i(\omega)\to0$ as
$k_b\to\infty$, at any finite value of $\omega$, but that
$\int_0^\infty \!\Delta P_i(\omega)\to 2\pi\Delta\sigma^2_i$.  The
reason is that, whilst $\Delta P_i(\omega)$ vanishes as $1/k_b$, it
extends over a frequency range $0<\omega\alt k_b$, so the total
integrated contribution does not vanish in the limit $k_b\to\infty$.
In words, the equilibration reaction contributes high-frequency noise
which vanishes at any particular finite frequency in the limit of
infinitely fast equilibration, but makes a net non-zero contribution
to the total integrated noise.

The universal magnitude of the sum rule deficit can be understood by
analysing the simplest of all equilibration reactions, namely
\begin{equation}
\text{A}\tofromrate{k_{\!f}}{k_b}\text{A}^*.\label{eq:AAs}
\end{equation}
Solving for the point statistics in this system one has $K\myav{N_A} =
\myav{N_{A^*}}$ and, from Eqs.~\eqref{eq:app:vcm} in the Appendix, 
\begin{equation}
\begin{array}{l}
-K\sigma_A^2+S_{AA^*}+\myav{N_{A^*}}=0\\[6pt]
-K S_{AA^*}+\sigma_{A^*}^2
+K\sigma_A^2-S_{AA^*}-2\myav{N_{A^*}}=0\\[6pt]
K S_{AA^*}-\sigma_{A^*}^2+\myav{N_{A^*}}=0
\end{array}
\label{eq:aasvcm}
\end{equation}
These equations are linearly dependent and cannot by themselves be
solved for the elements of the variance-covariance matrix.  The reason
is that the slow variable $N_A+N_{A^*}$ is exactly conserved by the
reaction scheme in Eq.~\eqref{eq:AAs}.  However this implies that
$\Delta N_A=-\Delta N_{A^*}$ and so $\sigma_A^2 = \sigma_{A^*}^2 =
-S_{AA^*}$.  These are consistent with Eqs.~\eqref{eq:aasvcm}, which
can now be solved to get
\begin{equation}
\sigma_A^2=\sigma_{A^*}^2=-S_{AA^*}=\frac{\myav{N_{A^*}}}{K+1}.
\end{equation}
This explains the universal value found for the sum rule deficit
above.

\begin{figure}
\begin{center}
\includegraphics{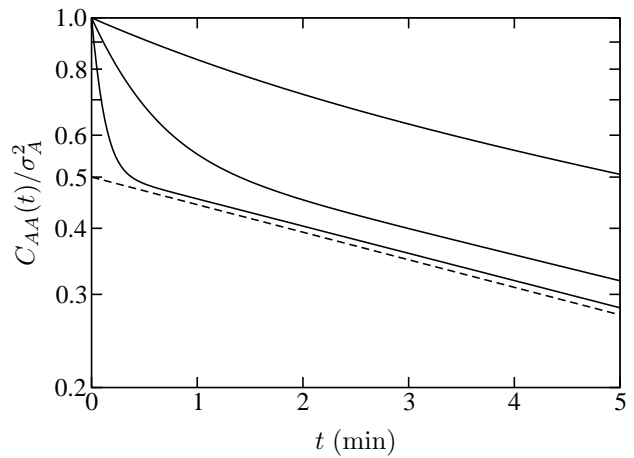}
\end{center}
\caption[?]{Correlation function for the model in Eq.~\eqref{eq:AAs2},
with $\gamma_R = K\gamma / (K+1) = 0.12\, \mathrm{min}^{-1}$ and $K =
1$.  The lines are for $k_b = 0.2$, 1.0, $5.0\,\mathrm{min}^{-1}$, and
$k_b\to\infty$ (dashed line).\label{fig:fig2}}
\end{figure}

To complete the discussion, we turn finally to the correlation
functions where the analogue of Eq.~\eqref{eq:limorder} is
\begin{equation}
\lim_{k_b\to\infty}\;\lim_{t\to0}\;C_{ii}(t) \ne
\lim_{t\to0}\;\lim_{k_b\to\infty}C_{ii}(t)
\end{equation}
for $i=\text{A}$ and $\text{A}^*$.  The left hand side is always equal
to $\sigma_i^2$.  One can show that the difference between the two
sides gives an alternative expression for the sum rule deficit,
\begin{equation}
\Delta\sigma_i^2 = \sigma_i^2-\lim_{t\to0} C_{ii}(t).
\end{equation}
This can be made clear by another example.  Consider
\begin{equation}
\torate{k}\text{A}\tofromrate{k_{\!f}}{k_b}\text{A}^*\torate{\gamma}.
\label{eq:AAs2}
\end{equation}
The steady state point statistics, which hold for arbitrary $k_b$, are
$\myav{N_A} = (k_b+\gamma)\myav{N_{A^*}}/k_f$,
$\myav{N_{A^*}} = k/\gamma$, $\sigma_A^2 / \myav{N_A} =
\sigma_{A^*}^2 / \myav{N_{A^*}} = 1$, $S_{AA^*}=0$.  We solve for the
correlation functions using the methods described in section \ref{sec:gen}.
Similar results are obtained for both A and $\text{A}^*$ so we report
results only for species A.  The Laplace-transformed autocorrelation
function is
\begin{equation}
\tilde C_{AA}(s) = 
\frac{\sigma_A^2(s+\gamma+k_b)}
{s^2+(\gamma+k_f+k_b)s+\gamma k_f}.
\end{equation}
Writing
\begin{equation}
s^2+(\gamma+k_f+k_b)s+\gamma k_f = (s+k_+)(s+k_-)
\end{equation}
allows us to perform the inverse Laplace transform by the method of
partial fractions.  The decay rates $k_\pm$ are given by
\begin{equation}
k_\pm = \frac{(\gamma+k_f+k_b)\pm
\sqrt{(\gamma+k_f+k_b)^2-4\gamma k_f}}{2}.
\end{equation}
The net result is 
\begin{equation}
C_{AA}(t) = \sigma_A^2 ( Ae^{-k_+t}+Be^{-k_-t})
\end{equation}
with
\begin{equation}
A=\frac{(\gamma+k_b)-k_-}{k_+-k_-},\quad B=1-A.
\end{equation}
For any finite value of $k_b$, it is clear from this that
$C_{AA}(0)=\sigma_A^2$ and there is no sum rule deficit.

Now let us take the limit $k_b\to\infty$ with $K=k_f/k_b$ being held
constant.  To leading order we have
\begin{equation}
\begin{array}{ll}
\displaystyle
k_+\to(K+1)\,k_b,\quad{}
&\displaystyle
k_-\to\frac{K\gamma}{K+1},\\[12pt]
\displaystyle
A\to\frac{K}{K+1},
&\displaystyle
B\to\frac{1}{K+1},
\end{array}
\end{equation}
and thus
\begin{equation}
\lim_{k_b\to\infty}
C_{AA}(t)
= \frac{\sigma_A^2}{K+1}\,
\exp\Bigl[-\frac{K\gamma t}{K+1}\Bigr].
\end{equation}
Apart from the amplitude, this correlation function is characteristic
of $\torate{k}\text{X}\torate{\gamma_R}$ with $\gamma_R = K\gamma /
(K+1)$.  This is entirely in accord with rewriting the original scheme
in terms of the relevant slow variable.

The analysis shows that in the limit $k_b\to\infty$, we have
$C_{AA}(0)<\sigma_A^2$.  Thus there is a non-zero sum rule deficit
\begin{equation}
\Delta\sigma_A^2=\sigma_A^2\Bigl(1-\frac{1}{K+1}\Bigr)=
\frac{\myav{N_{A^*}}}{K+1}
\end{equation}
where we have used $\sigma_A^2 = \myav{N_A} = \myav{N_{A^*}}/K$.
Again, the universal value for the sum rule deficit is recovered.

Our analysis shows what happens to the correlation functions in the
limit of fast equilibration (similar results are found for
$C_{A^*\!A^*}$).  At any finite $k_b$ there are two decay rates in the
correlation function.  In the limit $k_b\to\infty$ one of these decay
rates diverges whilst the other saturates to a finite value.  The
diverging decay rate carries with it a finite amplitude and it is this
that gives rise to the sum rule deficit.  The phenomenology is
illustrated in Fig.~\ref{fig:fig2} which shows typical results for
$C_{AA}(t)$ at several increasing values of the equilibration rate.

The results in this section are supportive of the general strategy of
elimination of the fast variables in terms of the appropriate slow
variables.  To be specific, the strategy is to rewrite the chemical
rate equations in terms of fast and slow variables, eliminate the fast
variables, and re-interpret the reduced rate equations in terms of a
reduced reaction network, possibly involving new chemical species (in
the present example, X replaces A and A$^*$).  Our study has provided
examples where one can rigorously prove that this strategy is
successful.

\section{Discussion}
Firstly, let us make some general remarks about the use of noise power
spectra to characterise the stochastic properties of chemical reaction
networks.  As we have shown, the noise power spectra are
straightforward to calculate once one has the point statistics (the
mean values and the variance-covariance matrix), for a linear network,
or for a linearisation of a non-linear network.  In fact, the power
spectra are \emph{easier} to calculate than the autocorrelation
functions, which in general require the factorisation of a polynomial,
unless one is satisfied with stopping at the Laplace-transformed
autocorrelation functions.  Because the power spectra are functions of
frequency, they are a more refined measure than the point statistics,
and can be used for instance as a sensitive test of whether two
reaction networks can be mapped onto one another.

Let us comment briefly on the situation for experiments and
simulations.  Both face similar difficulties in measuring the power
spectra.  A signal of sufficiently long duration needs to be captured
at sufficiently fine resolution in order to be able to estimate the
power spectrum.  Event-driven simulations though, such as those based
on the Gillespie algorithm \cite{GillKMC}, have an advantage since in
principle the exact moments when a signal changes its value are known.

Secondly, we discuss the biological relevance of our results.  Our
results indicate that the way noise is transmitted through a signal
detection motif may be more complicated than previously thought
\cite{Paulsson, Shibata, TNWtW}.  In particular, if the detection
motif consumes the input signal molecules, a correlation can be set up
between the intrinsic noise generated by the detection motif, and the
input signal noise.  We have examined this in the context of a
post-translational modification reaction attached to the output of a
gene expression module.  Our analysis shows incidentally that
post-translational modification can ameliorate the noise of gene
expression, by smoothing out the `burstiness' of the translation step.

We also examined the effect of a fast equilibration reaction
interposed in the network.  We find that, in the limit of infinitely
fast equilibration, such a network can be exactly mapped onto a
reduced or coarse-grained network through the use of suitably chosen
slow variables.  This result supports the use of slow variables as a
general strategy for model-order reduction by elimination of fast
reactions.

This work was supported by the Amsterdam Centre for Computational
Science (ACCS).  The work is part of the research programme of the
``Stichting voor Fundamenteel Onderzoek der Materie (FOM)'', which is
financially supported by the ``Nederlandse organisatie voor
Wetenschappelijk Onderzoek (NWO)''.  PBW acknowledges the hospitality
of Trey Ideker and the California Institute for Telecommunications and
Information Technology (Calit2) at the University of California, San
Diego, where part of this work was written up.


\appendix

\section{Technical details}
Technical details and proofs are relegated to this Appendix.  For
further reading we recommend Gardiner \cite{gardinerbook}, van Kampen
\cite{kampenbook}, Gillespie \cite{GillCLE, GillKMC}, and Swain
\cite{Swain}.
\subsection{Point statistics}
\label{sec:appkb}
We first describe the chemical master equation and present a useful
moment relation.  As a preliminary step we need to define some basic
quantities.  The stoichiometry matrix is a non-square matrix
$\nu_{i\alpha}$ equal to the change in species $i$ due to the firing
of reaction $\alpha$ in the network.  We shall use roman indices to
label chemical species and the greek symbol $\alpha$ to label
reactions.  In terms of $\nu_{i\alpha}$, the chemical rate equations
can be written
\begin{equation}
\frac{d N_i}{dt}={\textstyle\sum_\alpha} \nu_{i\alpha} a_\alpha
\label{eq:app:cre}
\end{equation}
where $a_\alpha$ is the flux through reaction channel $\alpha$.  The quantity
$a_\alpha$ is usually dependent on the current values of $N_i$ and is
referred to by Gillespie as the `propensity function'.  The propensity
functions for a linear reaction network will be described shortly.

In these terms, the probability
$P_{N_i}$ of being in the state characterised by $N_i$ molecules of
species $i$ obeys
\begin{equation}
\frac{\partial P_{N_i}}{\partial t} =
\sum_\alpha
\Bigl\{a_\alpha(N_i-\nu_{i\alpha})\,P_{N_i-\nu_{i\alpha}}
-a_\alpha(N_i)\,P_{N_i}\Bigr\}.\label{eq:app:cme}
\end{equation}
If a reaction channel is impossible, for instance where
$N_i-\nu_{i\alpha}<0$, we set $a_\alpha = 0$.  Eq.~\eqref{eq:app:cme}
is the chemical master equation.

Multiplying Eq.~\eqref{eq:app:cme} by a general function $f(N_i)$ and
summing over all $N_i$ yields
\begin{equation}
\frac{d\myav{f}_t}{dt} =
{\textstyle\sum_\alpha}
\myav{a_\alpha(N_i)\,[f(N_i+\nu_{i\alpha})-f(N_i)]}_t.
\label{eq:app:mcme}
\end{equation}
This is particularly useful for deriving equations for moments.

We now indicate how the point statistics and in particular the
variance-covariance matrix can be found for an arbitrary linear
reaction network.  The results follow as particular cases of
Eq.~\eqref{eq:app:mcme}.

First we need to specify in more detail the propensity functions for a
linear reaction network.  We distinguish between zeroth- and
first-order reactions in the network.  The former are pure generating
reactions of the form `$\to\text{products}$'.  The latter are
monomolecular reactions and comprise the rest of the network (in a
linear reaction network, all reactions are either zeroth- or
first-order).  The set of zeroth-order reactions is denoted by $\zero$
and the set of first-order reactions by $\one$.

The propensity function for a zeroth-order reaction is
\begin{equation}
a_\alpha=k_\alpha,\quad \alpha\in\zero.\label{eq:app:v0}
\end{equation}
where $k_\alpha$ is the reaction rate.  For the first order reactions,
we introduce an indicator matrix $\epsilon_{i\alpha}$, such that
$\epsilon_{i\alpha} = 1$ if the reaction $\alpha$ involves species $i$
as the reactant, and $\epsilon_{i\alpha} = 0$ otherwise.  Armed with
this, the propensity function for a first-order reaction is
\begin{equation}
a_\alpha=k_\alpha{\textstyle\sum_i}\epsilon_{i\alpha}N_i
\quad \alpha\in\one.\label{eq:app:v1}
\end{equation}
Inserting Eqs.~\eqref{eq:app:v0} and \eqref{eq:app:v1} into
Eq.~\eqref{eq:app:cre} obtains Eq.~\eqref{eq:cre} in the main text,
with
\begin{equation}
b_i=\sumzero k_\alpha \nu_{i\alpha}\,,\quad
K_{ij}=\sumone k_\alpha \nu_{i\alpha}\epsilon_{j\alpha}\,.
\end{equation}

The first moment of the chemical master equation is found by setting
$f(N_i)=N_i$ in Eq.~\eqref{eq:app:mcme}.  It is easy to show that this
gives Eq.~\eqref{eq:nit} in the main text, with the above definitions
of $b_i$ and $K_{ij}$.

The second moment of the chemical master equation is found by setting
$f(N_i)=N_iN_j$ in Eq.~\eqref{eq:app:mcme}.  This gives a closed set
of equations for the elements of the variance-covariance matrix.  In
steady state, these equations can be reduced to
\begin{equation}
{\textstyle\sum_k}(K_{ik} S_{kj}+K_{jk} S_{ki})+H_{ij}=0
\label{eq:app:vcm}
\end{equation}
where
\begin{equation}
\textstyle
H_{ij}=\sumzero k_\alpha \nu_{i\alpha} \nu_{j\alpha}+\sum_k\Bigl(
\sumone k_\alpha \nu_{i\alpha} \nu_{j\alpha} \epsilon_{k\alpha}
\Bigr)\myav{N_k}
\label{eq:app:h}
\end{equation}
Eq.~\eqref{eq:app:vcm} is a set of linear simultaneous equations for
the elements $S_{ij}$ of the variance-covariance matrix.

As an example of the application of this machinery, consider the model
for gene expression in Eq.~\eqref{eq:MA} in the main text.  There are
two species, M and A, and four reactions.  The stoichiometry matrix is
\begin{equation}
\nu_{i\alpha}=\Bigl(
\begin{array}{c}1\\0\end{array}\Big|
\begin{array}{ccc}-1&0&0\\0&1&-1\end{array}\Bigr).
\end{equation}
The reaction rates are
\begin{equation}
k_\alpha=(\;k\;|\;\lambda\;\;\rho\;\;\gamma\;).
\end{equation}
The indicator matrix is
\begin{equation}
\epsilon_{i\alpha}=\Bigl(
\begin{array}{c}\cdot\\\cdot\end{array}\Big|
\begin{array}{ccc}1&1&0\\0&0&1\end{array}\Bigr).
\end{equation}
In these, the reactions to the left of the vertical line correspond to 
$\alpha\in\zero$ and the reactions to the right of the line correspond to
$\alpha\in\one$ (the indicator matrix is only defined for the latter
reactions). 

From these one computes
\begin{equation}
K_{ij}=\Bigl(\begin{array}{cc}
-\lambda & 0\\ \rho & -\gamma \end{array}\Bigr).
\end{equation}
Obviously this could have been written down by inspection. Less
trivially, one also finds
\begin{equation}
\begin{array}{ll}
\displaystyle
H_{ij}&=
\Bigl(\begin{array}{cc}k+\lambda\myav{N_M} & 0\\
0 & \rho\myav{N_M}+\gamma\myav{N_A}\end{array}\Bigr)\\[18pt]
\displaystyle
&=\Bigl(\begin{array}{cc}2\lambda\myav{N_M} & 0\\
0 & 2\gamma\myav{N_A}\end{array}\Bigr).
\end{array}
\end{equation}
Eqs.~\eqref{eq:app:vcm} simplify to
\begin{equation}
\begin{array}{l}
-2\lambda\sigma_M^2+2\lambda\myav{N_M}=0,\\[6pt]
\rho\sigma_M^2-\gamma S_{M\!A}-\lambda S_{M\!A} = 0,\\[6pt]
2\rho S_{M\!A}-2\gamma\sigma_A^2+2\gamma\myav{N_A}=0.
\end{array}
\end{equation}
There are only three equations since Eq.~\eqref{eq:app:vcm} is
symmetric in $i$ and $j$.  These are solved to get
Eqs.~\eqref{eq:MAvcm} in the main text.

\subsection{Regression theorem}
\label{sec:appc}
The proof of the regression theorem Eq.~\eqref{eq:corr} in the main
text is straightforward and parallels the development in Gardiner
\cite{gardinerbook}.  We start by writing out an explicit expression for
the correlation function
\begin{equation}
\begin{array}{l}
\displaystyle
C_{ij}(t)=\int d\{N_{i,0}\}\,d\{N_{i,t}\}\;
\Delta N_{i,0} \;\Delta N_{j,t}\\[6pt]
{}\hspace{8em}{}\times P_s(N_{i,0})\;P(N_{i,t}|N_{i,0}; t).
\end{array}
\label{eq:app:ceq}
\end{equation}
In this, $P_s(N_{i,0})$ is the steady state probability distribution
for the starting point $N_i = N_{i,0}$, and $P(N_{i,t}|N_{i,0}; t)$ is
the conditional probability distribution for $N_i = N_{i,t}$ at time
$t$, given that the system started with $N_i = N_{i,0}$ at $t = 0$.
These probability distributions could in principle be found by solving
the chemical master equation.  We now define two kinds of averages
\begin{equation}
\begin{array}{l}
\myav{f(N_i)}_{t|0}=
\int d\{N_{i,t}\}\;f(N_{i,t})\;P(N_{i,t}|N_{i,0}; t),\\[6pt]
\myav{f(N_i)}_0=
\int d\{N_{i,0}\}\;f(N_{i,0})\;P_s(N_{i,0}).
\end{array}
\end{equation}
In words, the first is the average value of a function of $N_i =
N_{i,t}$ at time $t$, given that the system started with $N_i =
N_{i,0}$ at $t = 0$.  The second is the average value of a function of
$N_i = N_{i,0}$ in steady state conditions.  In terms of these,
Eq.~\eqref{eq:app:ceq} can be written as
\begin{equation}
C_{ij}(t)=\myav{{}\,\Delta N_i\,\myav{\Delta N_j}_{t|0}\,}_0.
\label{eq:app:cgen}
\end{equation}
This result is completely general \cite{gardinerbook}.  

For a linear reaction network, taking the first moment of the chemical
master equation proves an analogue to Eq.~\eqref{eq:dnit}, namely
\begin{equation}
\frac{d\myav{\Delta N_i}_{t|0}}{dt} =
{\textstyle\sum_j}K_{ij}\myav{\Delta N_j}_{t|0}.
\end{equation}
Combining this with Eq.~\eqref{eq:app:cgen} demonstrates that
$C_{ij}(t)$ obeys Eq.~\eqref{eq:corr} in the main text.

\subsection{Chemical Langevin equations}
\label{sec:appl}
We prove that the power spectra obtained from the chemical Langevin
equations for a linear reaction network are equivalent to the exact
results obtained from the chemical master equation.  As the first
step, Gillespie has shown that the chemical Langevin equations for a
general network are \cite{GillCLE}
\begin{equation}
\frac{d N_i}{dt}={\textstyle\sum_\alpha} \nu_{i\alpha} a_\alpha
+{\textstyle\sum_\alpha} \nu_{i\alpha} a_\alpha^{1/2}\,\Gamma_\alpha.
\label{eq:app:cle}
\end{equation}
In this, $\Gamma_\alpha$ are independent unit Gaussian white noise
functions, one for each reaction channel.  Applying this to a linear
reaction network obtains
\begin{equation}
\frac{d N_i}{dt}={\textstyle\sum_j} K_{ij} N_j + b_i + \eta_i
\label{eq:app:cle2}
\end{equation}
with
\begin{equation}
\textstyle
\eta_i=\sumzero k_\alpha^{1/2} \nu_{i\alpha} \Gamma_\alpha
+\sumone 
(k_\alpha \sum_j\epsilon_{j\alpha}N_j)^{1/2}
\nu_{i\alpha} \Gamma_\alpha.
\end{equation}
It follows that the $\eta_i$ are Gaussian white noise functions with
the following statistics
\begin{equation}
\myav{\eta_i}=0,\quad
\myav{\eta_i(t)\,\eta_j(t')}=H_{ij}\,\delta(t-t').
\end{equation}
The correlation matrix $H_{ij}$ is \emph{identical} to the matrix that
features in the computation of the variance-covariance matrix, given
by Eq.~\eqref{eq:app:h} above, which can be interpreted in this
context as a fluctuation-dissipation theorem.

Working in terms of the deviation from steady state, and taking the
Fourier transform of the chemical Langevin equations, obtains
\begin{equation}
i\omega\,\Delta  N_i = 
{\textstyle\sum_j} K_{ij} \Delta  N_j +  \eta_i.
\label{eq:app:cle3}
\end{equation}
We conclude
\begin{equation}
P_i(\omega) = \myav{|\Delta N_i|^2} = 
{\textstyle\sum_{jk}}B_{ij} H_{jk} B_{ki}^\dag
\label{eq:app:picle}
\end{equation}
where $B_{ij}(i\omega)$ is the inverse of $i\omega\delta_{ij}-K_{ij}$
and $B_{ij}^\dag(i\omega) = B_{ji}(-i\omega)$ is the adjoint.  Our
task is to prove that Eq.~\eqref{eq:app:picle} is equivalent to
Eqs.~\eqref{eq:lcorr} and \eqref{eq:l2p} in the main text.

The problem is made easier if we rewrite everything in abstract matrix
notation.  The result we have just obtained can be written as
\begin{equation}
\mymat{P}_1=\mymat{B}\cdot \mymat{H}\cdot\mymat{B}^\dag
\label{eq:app:bhb}
\end{equation}
with
\begin{equation}
\mymat{B}=(i\omega\mymat{I}-\mymat{K})^{-1},\quad
\mymat{B}^\dag=(-i\omega\mymat{I}-\mymat{K}^\mathrm{T})^{-1}
\label{eq:app:b}
\end{equation}
($\mymat{K}^\mathrm{T}$ is the transpose of $\mymat{K}$).  The
diagonal elements of $\mymat{P}_1$ are the power spectra from the
chemical Langevin equation route.

We now similarly rewrite the results from the analysis of the chemical
master equation.  Eqs.~\eqref{eq:app:vcm}, \eqref{eq:lcorr} and
\eqref{eq:l2p} can be written respectively as
\begin{equation}
\begin{array}{l}
\mymat{K}\cdot\mymat{S}+\mymat{S}\cdot\mymat{K}^\mathrm{T}+\mymat{H}=0,\\[6pt]
i\omega \mymat{C}-\mymat{S}=\mymat{K}\cdot\mymat{C},
\quad\text{or}\quad\mymat{C} = \mymat{B}\cdot\mymat{S},\\[6pt]
\mymat{P}_2=\mymat{C}+\mymat{C}^\dag
=\mymat{B}\cdot\mymat{S} + \mymat{S}\cdot\mymat{B}^\dag
\end{array}
\label{eq:app:p2}
\end{equation}
(note that $\mymat{S}$ is real and symmetric).  The diagonal elements
of $\mymat{P}_2$ are the power spectra from the chemical master
equation route.

Eliminate $\mymat{H}$ between the first of Eqs.~\eqref{eq:app:p2} and
Eq.~\eqref{eq:app:bhb} to get
\begin{equation}
\mymat{P}_1=-(\mymat{B}\cdot\mymat{S}
\cdot\mymat{K}^\mathrm{T}\!\cdot\mymat{B}^\dag
+\mymat{B}\cdot\mymat{K}\cdot\mymat{S}\cdot\mymat{B}^\dag).
\label{eq:app:p1a}
\end{equation}
It follows from Eq.~\eqref{eq:app:b} that
\begin{equation}
\mymat{B}\cdot\mymat{K}=i\omega\mymat{B}-\mymat{I},\quad
\mymat{K}^\mathrm{T}\!\cdot\mymat{B}^\dag
=-i\omega\mymat{B}^\dag-\mymat{I}.
\end{equation}
On substituting these into Eq.~\eqref{eq:app:p1a}, there is a
cancellation of terms, and one finds
\begin{equation}
\mymat{P}_1=\mymat{B}\cdot\mymat{S}+\mymat{S}\cdot\mymat{B}^\dag.
\end{equation}
Comparison with the last of Eqs.~\eqref{eq:app:p2} shows that
$\mymat{P}_1=\mymat{P}_2$.  This contains our desired result, and
proves that the power spectra obtained by the chemical Langevin
equation route are the same as those obtained by the (exact) chemical
master equation route.  Actually, we have a slightly stronger result,
since it follows that $\myav{|\Delta N_X|^2} =
\mymat{x}^\mathrm{T}\!\cdot \mymat{P}_1\cdot\mymat{x} =
\mymat{x}^\mathrm{T}\!\cdot \mymat{P}_2\cdot\mymat{x}$, for any linear
combination $N_X=\sum_i x_i N_I$.

Swain has presented an analysis of the chemical Langevin equations for,
effectively, a general linear reaction network \cite{Swain}.  He
obtains general results for the correlation functions and the
variance-covariance matrix in terms of the eigenvalues of $K_{ij}$.
The present section can be read as proving that the
variance-covariance matrix can be found by a simpler route, by solving
Eq.~\eqref{eq:app:vcm} in terms of the noise correlation matrix.  For
the correlation functions though, our route involves an inverse
Laplace transform which in general requires that a polynomial be
factorised, and is equivalent to solving the secular equation for
$K_{ij}$.  So our methods do not provide any additional simplification
compared to Swain for the correlation functions.  As we have
emphasised in the main text though, our results demonstrate that the
power spectra are much easier to calculate than the correlation
functions.


\end{document}